\documentclass[12pt,epsfig]{article}
\usepackage{epsfig}
\begin{document}

\begin{center} 
{\Large {\bf The Scalar Potential of 
Supersymmetric 
$SU(3)_{C}\times SU(3)_{L}\times U(1)_{N}$ Model}}
\end{center}

\begin{center}
M. C. Rodriguez  \\
{\it Grupo de F{\'{\i}}sica Te\'{o}rica e Matem\'{a}tica F\'{\i}sica \\
Departamento de F\'{\i}sica  \\
Universidade Federal Rural do Rio de Janeiro (UFRRJ) \\
BR 465 Km 7, 23890-000 Serop\'{e}dica - RJ \\
Brasil}
\end{center}

\date{\today}

\begin{abstract}
We will study in details the full scalar 
potential of the Minimal Supersymmetric 
$SU(3)_{C}\times SU(3)_{L}\times U(1)_{N}$ Model. 
We will present numerical predictions for all the usual scalars of this 
model, we will show that the mass of the light scalar of CP par is 
$125.5$ GeV, such that its masses satisfy the current experimental 
constraints. 
\end{abstract}

PACS number(s): 12.60. Jv



\section{Introduction}

Models with the gauge symmetry 
\begin{equation}
SU(3)_{C} \times SU(3)_{L} \times U(1)_{N}
\end{equation} 
are known as $3-3-1$ for short. They are interesting  
possibilities for the physics at the TeV scale \cite{singer,ppf,331rh,Pleitez:1994pu,Ponce:2001jn}. It is a subgroup of 
unification group $E_{6}$ \cite{Sanchez:2001ua} and it is also 
an $SU(6) \times U(1)_{X}$ subgroup \cite{Martinez:2001mu}.

In fact, this may be the last symmetry involving the lightest elementary 
particles: leptons, and there are several distinct possible models based 
on this gauge symmetry. The reason for this is that the electric charge 
operator, in the $SU(3)_{L}$ generators, is defined as
\begin{eqnarray}
\frac{Q}{e}&=& \frac{1}{2}(\lambda_{3}- \vartheta \lambda_{8})+N \,\ I_{3 \times 3}, 
\label{co}
\end{eqnarray}
where the $\vartheta$ and $N$ are parameters defining differents representation contents and $\lambda_{3}$, $\lambda_{8}$ are the diagonal generators of $SU(3)$ given by
\begin{eqnarray}
\lambda_{3}= \left( \begin{array}{ccc} 
+1 & 0 & 0\\ 
0 & -1 & 0 \\
0 & 0 & 0          
\end{array} \right), \,\
\lambda_{8}= \frac{1}{\sqrt{3}} \left( \begin{array}{ccc} 
+1 & 0 & 0\\ 
0 & +1 & 0 \\
0 & 0 & -2          
\end{array} \right).
\end{eqnarray}

One of the possible and well-studied models in the literature is the 
model proposed by Pleitez-Pisano~\cite{ppf}, where we chose 
$\vartheta=\sqrt3$. In this case, Eq.(\ref{co}) become
\begin{eqnarray}
\frac{Q}{e}&=& \left( \begin{array}{ccc} 
N & 0 & 0 \\ 
0 & N-1 & 0 \\
0 & 0 & N+1          
\end{array} \right).
\label{mppf}
\end{eqnarray} 
The lepton sector is exactly the same as in the Standard Model (SM) 
\cite{sg} but now there is a symmetry, at large energies among, say 
$e^{-}$, $\nu_{e}$ and $e^{+}$ \cite{ppf} and its supersymmetric version has 
already been considered in 
Refs.~\cite{ema1,pal2,331susy1,mcr,Rodriguez:2005jt}, and we will call it as 
MSUSY331. There are also another 
possibility where the triplet $(\nu_{l_{i}}, l_{i}, L^{c}_{i})_{L}$ where $L$ 
is an extra charged leptons wchich do not mix with the known 
leptons \cite{Pleitez:1992xh,Pleitez:1993gc}. We want to remind that there is no right-handed 
(RH) neutrino in both  model presented above. The 3-3-1 model of 
Refs.~\cite{331rh}, is know as 3-3-1 model with right-handed neutrinos. we 
define $\vartheta=(1/\sqrt{3})$ and we get
\begin{eqnarray}
\frac{Q}{e}&=& \left( \begin{array}{ccc} 
N+ \frac{1}{3} & 0 & 0 \\ 
0 & N- \frac{2}{3} & 0 \\
0 & 0 & N+ \frac{1}{3}          
\end{array} \right),
\label{331rhn}
\end{eqnarray} 
and in the lepton sector, we have $e^{-}$, $\nu_{e}$ and 
$\nu^{c}_{e}$ \cite{331rh}, the supersymmetric version of this model was 
built in \cite{331susy2,huong}. Clearly, we can also have models similar to 
the two districts above, having heavy leptons, such as $E^{+}$ instead of $e^{+}$ or even $N^{c}$ replacing $\nu^{c}$, for more details see 
\cite{Pleitez:1994pu}.  

Once this symmetry is imposed on the lightest generation and extended to 
the other leptonic generations it follows that the quark sector must be 
enlarged by considering heavy exotic charged quarks, being able to have 
exotics with different charges than the current quarks, as well as having 
the possibility of quarks with the same charges but heavier than the current ones. 

Although this model coincides at low energies with the SM it explains 
some fundamental questions that are accommodated, but not explained, in the SM. These questions are:
\begin{enumerate}
\item The family number must be a multiple of three in order to cancel 
anomalies~\cite{ppf,331rh};
\item Why $\sin^{2} \theta_{W}<\frac{1}{4}$ is observed;
\item The models have a scalar sector similar to the two Higgs doublets Model;
\item It is the simplest model that includes bileptons of both types: 
scalars and vectors ones;
\item It solves the strong $CP$-problem;
\item The model has several sources of CP violation \cite{oravinez,Montero:2005yb}.
\end{enumerate}
The 3-3-1 models havebeen studied extensively over the last decade.

Recently it was reported a light Higgs boson at the LHC and its value is given by\cite{Aad:2015zhl}
\begin{equation}
M_{H} = 125.09 \pm 0.21\,\mathrm{(stat.)} \pm 0.11\,\mathrm{(syst.)}~\mathrm{GeV}.
\label{higgsexpvalue}
\end{equation}
The scalar sector of the 3-3-1 model was studied on 
\cite{Pleitez:1993gc,oravinez,tonasse}.  The scalar potential in several 
3-3-1 models was considered in \cite{Diaz:2003dk,Nguyen:1998ui,Ponce:2002sg,Giraldo:2011gd,
CarcamoHernandez:2014mlk}.

When we have presented the MSUSY331, we have shown the lighest higgs has 
mass around 124.5 GeV \cite{331susy1}, later we did the analysis of all 
the usual scalar spectrum  considering only triplets 
\cite{Rodriguez:2005jt} and we have also studied the mass spectrum 
at \cite{Rodriguez:2010tn}. Our goal is to extend, in a similar way to what 
was done in \cite{Rodriguez:2005jt}, when we consider the sextet 
and anti-sextet and in this case we have shown we can explaing recent 
experimental data regarding the $W$-boson mass presented by the CDF 
\cite{Rodriguez:2022hsj}.

This paper is organized as follows. In Sec. \ref{msusy1} we review the 
minimal Supersymmetric 3-3-1 Model with $R$-Parity Conservation. Our resutls for the scalar potential 
with the anti-Sextet is presented in Sec. \ref{sec:sp}. 
Finally, the last section is devoted to our 
conclusions. In App.(\ref{psusy1}) we review the scalar potential of 
3-3-1 Model without Supersymmetry, while in App.(\ref{constpot}) we 
show how to get an analytical expression to our scalar potential in 
order to compare its expresion with the ones used at 3-3-1 Model.

\section{Minimal Supersymmetric 3-3-1 model.}
\label{msusy1}

This supersymmetric version of the 3-3-1 model given at \cite{ppf} was 
presented at \cite{ema1,pal2,331susy1,mcr,Rodriguez:2010tn}. All 
the superfields of this model we briefly presented in a recent 
study \cite{Rodriguez:2022hsj}.

In the minimal 331 model, denoted as m331, the scalar sector is composed 
of  three scalar triplets:
\begin{eqnarray}
\eta &=& 
\left( \begin{array}{c} 
\eta^{0} \\ 
\eta^{-}_{1} \\
\eta^{+}_{2}          
\end{array} \right) 
\sim ({\bf 1},{\bf 3},0),\quad
\rho = 
\left( \begin{array}{c} 
\rho^{+} \\ 
\rho^{0} \\
\rho^{++}          
\end{array} \right) 
\sim ({\bf 1},{\bf 3},+1), \nonumber \\ 
\chi &=& 
\left( \begin{array}{c} 
\chi^{-} \\ 
\chi^{--} \\
\chi^{0}          
\end{array} \right) 
\sim ({\bf 1},{\bf 3},-1).
\label{3t} 
\end{eqnarray}
In parenthesis it appears the transformations properties under the respective
factors $(SU(3)_{C},SU(3)_{L},U(1)_{N})$. Those scalars, defined at 
Eq.(\ref{3t}), give masses 
for all the quarks and the charged leptons are massless. In order to 
give masses for the charged leptons, at tree level, we have to introduce 
the following anti-sextet
\begin{equation}
S = \left( \begin{array}{ccc} 
\sigma^{0}_{1}& 
\frac{h^{+}_{2}}{ \sqrt{2}}& \frac{h^{-}_{1}}{ \sqrt{2}} \\ 
\frac{h^{+}_{2}}{ \sqrt{2}}& H^{++}_{1}& \frac{ \sigma^{0}_{2}}{ \sqrt{2}} \\
 \frac{h^{-}_{1}}{ \sqrt{2}}& 
\frac{\sigma^{0}_{2}}{ \sqrt{2}}&  H^{--}_{2}        
\end{array} \right) \sim ({\bf1},{\bf6}^{*},0) \,\ . 
\label{sextet}
\end{equation}

The scalars get the following vacuum expectation values (VEVs) 
\cite{331susy1,mcr}:
When we break the 331 symmetry to the $SU(3)_{C} \otimes U(1)_{EM}$, the scalars get the following vacuum expectation values (VEVs):
\begin{equation} 
< \eta > = 
      \left( \begin{array}{c} v \\ 
                  0 \\
                  0          \end{array} \right),\quad 
< \rho > = 
      \left( \begin{array}{c} 0 \\ 
                  u \\
                  0          \end{array} \right),\quad 
< \chi > = 
      \left( \begin{array}{c} 0 \\ 
                  0 \\
                  w          \end{array} \right),
\label{vev} 
\end{equation}
and
\begin{equation}
< S > = 
      \left( \begin{array}{ccc} 
0 & 0& 0 \\ 
0& 0& \frac{z}{ \sqrt{2}} \\
0& \frac{z}{ \sqrt{2}}&  0
\end{array} \right),
\label{vevsemsusy} 
\end{equation}
where $v=v_{\eta}/ \sqrt{2}$, $u=v_{\rho}/ \sqrt{2}$, $w=v_{\chi}/ 
\sqrt{2}$ and $z=v_{\sigma_2}/ \sqrt{2}$. In a future analyses we want 
to allow that the field $\sigma_1$ get VEV as we have done 
\cite{Rodriguez:2022hsj}.

\begin{eqnarray} 
v&=&\frac{v_{\eta}}{\sqrt{2}}
\left( 1+ 
\frac{H_{\eta}+\imath F_{\eta}}{|v_{\eta}|}  
\right), \quad 
u= \frac{v_{\rho}}{\sqrt{2}}
\left( 1+ 
\frac{H_{\rho}+\imath F_{\rho}}{|v_{\rho}|}  
\right), \nonumber \\
w&=&\frac{v_{\chi}}{\sqrt{2}}
\left( 1+ 
\frac{H_{\chi}+\imath F_{\chi}}{|v_{\chi}|}  
\right), \quad                   
z=\frac{v_{\sigma^{0}_{2}}}{\sqrt{2}}
\left( 1+ 
\frac{H_{\sigma^{0}_{2}}+\imath F_{\sigma^{0}_{2}}}{|v_{\sigma^{0}_{2}}|}  
\right).
\label{vevantisextet} 
\end{eqnarray}

Besides, in order to to cancel chiral anomalies generated by the
superpartners of the scalars, we have to add the following higgsinos in the
respective anti-multiplets, 
\begin{eqnarray} 
\tilde{\eta}^{\prime} &=& 
\left( 
\begin{array}{c} 
\tilde{\eta}^{\prime0} \\ 
\tilde{\eta}^{\prime+}_{1} \\
\tilde{\eta}^{\prime-}_{2}          
\end{array} \right) 
\sim ({\bf1},{\bf3}^{*},0),\quad
\tilde{\rho}^{\prime} = 
\left( \begin{array}{c} 
\tilde{\rho}^{\prime-} \\ 
\tilde{\rho}^{\prime0} \\
\tilde{\rho}^{\prime--}          
\end{array} 
\right) 
\sim ({\bf1},{\bf3}^{*},-1), \nonumber \\ 
\tilde{\chi}^{\prime} &=& 
\left( \begin{array}{c} 
\tilde{\chi}^{\prime+} \\ 
\tilde{\chi}^{\prime++} \\
\tilde{\chi}^{\prime0}          
\end{array} \right) 
\sim ({\bf1},{\bf3}^{*},+1),
\label{shtc}  
\end{eqnarray}
\begin{equation}
\tilde{S}^{\prime} = 
      \left( \begin{array}{ccc} 
\tilde{\sigma}^{\prime0}_{1}& \frac{\tilde{h}^{\prime-}_{2}}{ \sqrt{2}}& 
\frac{\tilde{h}^{\prime+}_{1}}{ \sqrt{2}} \\ 
\frac{\tilde{h}^{\prime-}_{2}}{ \sqrt{2}}& \tilde{H}^{\prime--}_{1}& 
\frac{ \tilde{\sigma}^{ \prime 0}_{2}}{ \sqrt{2}} \\
 \frac{\tilde{h}^{\prime+}_{1}}{ \sqrt{2}}& 
\frac{ \tilde{\sigma}^{ \prime 0}_{2}}{ \sqrt{2}}&  
\tilde{H}^{\prime++}_{2}        
\end{array} \right)_L \sim ({\bf1},{\bf6},0).
\label{shsc} 
\end{equation}
There are also the scalar partners of the
Higgsinos defined in Eqs.~(\ref{shtc},\ref{shsc}) and we will denote them 
$\eta^{\prime}, \rho^{\prime}, \chi^{\prime},S^{\prime}$ \cite{331susy1,mcr}. This model was 
presented at \cite{ema1,pal2,331susy1,mcr}.

The usual scalars get the vacuum expectation values (VEVs) defined at 
Eqs.(\ref{vevsemsusy},\ref{vevantisextet}). In 
similar way for 
the new scalars we can write \cite{331susy1,mcr}:
\begin{eqnarray} 
v^{\prime}&=&
\frac{v_{\eta^{\prime}}}{\sqrt{2}}
\left( 1+ 
\frac{H_{\eta^{\prime}}+\imath 
F_{\eta^{\prime 0}}}{|v_{\eta^{\prime}}|}  
\right), \quad
u^{\prime}=\frac{v_{\rho^{\prime}}}{\sqrt{2}}
\left( 1+ 
\frac{H_{\rho^{\prime}}+\imath 
F_{\rho^{\prime}}}{|v_{\rho^{\prime}}|}  
\right),
\nonumber \\
w^{\prime}&=&\frac{v_{\chi^{\prime}}}{\sqrt{2}}
\left( 1+ 
\frac{H_{\chi^{\prime}}+\imath 
F_{\chi^{\prime}}}{|v_{\chi^{\prime}}|}  
\right), \quad 
z^{\prime}=
\frac{v_{\sigma^{\prime 0}_{2}}}{\sqrt{2}}
\left( 1+ 
\frac{H_{\sigma^{\prime 0}_{2}}+\imath F_{\sigma^{\prime 0}_{2}}}{|v_{\sigma^{\prime 0}_{2}}|}  
\right).
\label{vevantisextet} 
\end{eqnarray}
It is the complete set of fields in the minimal supersymmetric 3-3-1 model (MSUSY331). For those interested in the total Lagrangian of the model, 
we recommend seeing \cite{331susy1,mcr}.

\subsection{Genral Superpotential at MSUSY331}

The superpotential of our model is given by
\begin{equation}
W=W_{2}+W_{3}+ \bar{W}_{2}+ \bar{W}_{3}, 
\label{sp1}
\end{equation}
with $W_{2}$ having only two chiral superfields while $W_{3}$ has three chiral superfields. The terms allowed by  our symmetry are
\begin{eqnarray}
W_{2}&=&
\mu_{0i}(\hat{L}_{i}\hat{\eta}^{\prime})
+ 
\mu_{ \eta} 
(\hat{\eta}\hat{\eta}^{\prime})
+
\mu_{ \rho} 
(\hat{\rho}\hat{\rho}^{\prime})
+ 
\mu_{ \chi} 
(\hat{\chi}\hat{\chi}^{\prime})
+
\mu_{S} Tr[(\hat{S} \hat{S}^{\prime})], 
\label{w2geral}
\end{eqnarray}
and we also have
\begin{eqnarray}
W_{3}&=& \lambda_{1ijk} 
(\epsilon \hat{L}_{i}\hat{L}_{j}\hat{L}_{k})+
\lambda_{2ij} (\epsilon \hat{L}_{i}\hat{L}_{j}\hat{\eta})+
\lambda_{3ij} 
(\hat{L}_{i}\hat{S}\hat{L}_{j}) +
\lambda_{4i}(\epsilon\hat{L}_{i}\hat{\chi}\hat{\rho}) \nonumber \\
&+&
f_{1} 
(\epsilon \hat{\rho} \hat{\chi}\hat{\eta})
+
f_{2} 
(\hat{\chi}\hat{S}\hat{\rho})  
+
f_{3} 
(\hat{\eta}\hat{S}\hat{\eta}) +
f_{4}\epsilon_{ijk}\epsilon_{lmn}\hat{S}_{il}
\hat{S}_{jm}\hat{S}_{kn}+
f^{\prime}_{1} (\epsilon \hat{\rho}^{\prime}\hat{\chi}^{\prime} \hat{\eta}^{\prime})+
f^{\prime}_{2} (\hat{\chi}^{\prime} 
\hat{S}^{\prime}\hat{\rho}^{\prime})  
\nonumber \\ &+&
f^{\prime}_{3} (\hat{\eta}^{\prime} 
\hat{S}^{\prime}\hat{\eta}^{\prime}) 
+
f^{\prime}_{4}\epsilon_{ijk}\epsilon_{lmn}
\hat{S}^{\prime}_{il}\hat{S}^{\prime}_{jm}
\hat{S}^{\prime}_{kn} +
\kappa_{1i}
(\hat{Q}_{3} \hat{\eta}^{\prime}) \hat{u}^{c}_{i}+ 
\kappa_{2i} 
(\hat{Q}_{3} \hat{\rho}^{\prime}) \hat{d}^{c}_{i} +
\kappa_{3} 
(\hat{Q}_{3}\hat{\chi}^{\prime}) 
\hat{J}^{c} 
\nonumber \\
&+&
\kappa_{4\alpha i} 
(\hat{Q}_{\alpha} \hat{\eta}) \hat{d}^{c}_{i}+
\kappa_{5\alpha i} 
(\hat{Q}_{\alpha} \hat{\rho}) \hat{u}^{c}_{i} +
\kappa_{6\alpha\beta}
(\hat{Q}_{\alpha} \hat{\chi}) \hat{j}^{c}_{\beta} + 
\kappa_{7\alpha ij} 
(\hat{Q}_{\alpha} \hat{L}_{i}) 
\hat{d}^{c}_{j} \nonumber \\
&+&
\xi_{1ijk} \hat{d}^{c}_{i} \hat{d}^{c}_{j} \hat{u}^{c}_{k}
+
\sum_{\beta =1}^{2} \left(
\xi_{2ij\beta} \hat{u}^{c}_{i} \hat{u}^{c}_{j} \hat{j}^{c}_{\beta}+
\xi_{3 i\beta} \hat{d}^{c}_{i} \hat{J}^{c} \hat{j}^{c}_{\beta} \right) . 
\label{sp3m1}
\end{eqnarray}
The terms proportional for $f_{4}$ and $f^{\prime}_{4}$ were not 
introduced  in our previous works done at \cite{331susy1,mcr}.

\subsection{$R$-Parity charges at MSUSY331}

Some terms in the superpotential $W$ given in Eq.~(\ref{sp1}) violate the conservation of the new quantum number defined as 
\begin{equation}
{\cal F}=B+L
\end{equation}
quantum number, where $B$ is the baryon number while $L$ is the lepton number 
\cite{Pleitez:1992xh}. For instance, if we allow the 
$\xi_1$ term it implies 
in proton decay~\cite{pal2}. 

However, if we assume the global $U(1)_{\cal F}$ 
symmetry, it allows us to introduce the $R$-conserving symmetry, defined as 
\begin{equation}
R=(-1)^{3{\cal F}+2S}.
\end{equation}
The ${\cal F}$ number attribution is
\begin{equation}
\begin{array}{c}
{\cal F}(U^{--})={\cal F}(V^{-}) = - {\cal F}(J)= {\cal F}(j_{1,2})=
{\cal F}(\rho^{--})  \\  
= {\cal F}(\chi^{--}) ={\cal F}(\chi^{-}) = 
{\cal F}(\eta^{-}_{2})={\cal F}(\sigma_{1}^{0})=2,
\end{array}
\label{efe}
\end{equation}

As in the Minimal Supersymmetric Standard Model this definition implies that all known
standard model's particles have even $R$-parity while their supersymmetric
partners have odd $R$-parity. We can 
choose the following $R$- charges
Choosing the following R-charges \cite{Rodriguez:2010tn}  
\begin{eqnarray}
n_{\eta}&=&n_{\rho^{\prime}}=n_{S}=-1, \,\
n_{\rho}=n_{\eta^{\prime}}=
n_{S^{\prime}}=1, \,\
n_{\chi}=n_{\chi^{\prime}}=0, \nonumber \\
n_{L}&=&n_{Q_{i}}=n_{d_{i}}=1/2, \,\
n_{J_{i}}=-1/2, \,\ n_{u}=-3/2,
\label{rdiscsusy331} 
\end{eqnarray}
The terms  
which are proportional to the following constants: $\mu_{0}$;
$\lambda_{1},\lambda_{4},f_{3,4},
f^{\prime}_{3,4}, \kappa_{7},\xi_{1,2,3}$ violate 
the $R$-parity defined at Eq.(\ref{rdiscsusy331}).

The terms which satisfy the defined above symmetry (\ref{rdiscsusy331}) the 
term allowed by this $R$-charges in our superpotential, given at 
Eq.(\ref{sp3m1}), are given by
\begin{eqnarray} 
W_{RC}&=&W_{2RC}+W_{3RC},
\end{eqnarray}
where
\begin{eqnarray}
W_{2RC}&=&\mu_{ \eta} 
(\hat{\eta}
\hat{\eta}^{\prime})+
\mu_{ \rho} 
(\hat{\rho}
\hat{\rho}^{\prime})+
\mu_{ \chi} 
(\hat{\chi}
\hat{\chi}^{\prime})+ 
\mu_{S} Tr[(\hat{S} \hat{S}^{\prime})], \nonumber \\
W_{3RC}&=&
\lambda_{2ij} (\epsilon \hat{L}_{i}\hat{L}_{j}\hat{\eta})+
\lambda_{3ij} 
(\hat{L}_{i}\hat{S}
\hat{L}_{j})+
\lambda_{4i}(\epsilon\hat{L}_{i}
\hat{\chi}\hat{\rho})+
f_{1} 
(\epsilon \hat{\rho} \hat{\chi}\hat{\eta})+
f_{2} 
(\hat{\chi}\hat{S}\hat{\rho})  
\nonumber \\
&+&
f^{\prime}_{1} (\epsilon \hat{\rho}^{\prime}\hat{\chi}^{\prime} \hat{\eta}^{\prime})+
f^{\prime}_{2} (\hat{\chi}^{\prime} 
\hat{S}^{\prime}\hat{\rho}^{\prime})+
\kappa_{1i}
(\hat{Q}_{3} \hat{\eta}^{\prime}) \hat{u}^{c}_{i}+ 
\kappa_{2i} 
(\hat{Q}_{3} \hat{\rho}^{\prime}) \hat{d}^{c}_{i}
+
\kappa_{3} 
(\hat{Q}_{3}\hat{\chi}^{\prime}) 
\hat{J}^{c} 
\nonumber \\
&+&
\kappa_{4\alpha i} 
(\hat{Q}_{\alpha} \hat{\eta}) \hat{d}^{c}_{i}+
\kappa_{5\alpha i} 
(\hat{Q}_{\alpha} \hat{\rho}) \hat{u}^{c}_{i} 
+
\kappa_{6\alpha\beta}
(\hat{Q}_{\alpha} \hat{\chi}) \hat{j}^{c}_{\beta}.
\label{superpotentialrcmsusy331}
\end{eqnarray}

The soft terms to break SUSY\footnote{We are not considerating the Gaugino 
Mass terms neither sleptons and squarks, about the soft terms you can see \cite{331susy1}.} and conserve our $R$-Parity is written as
\begin{eqnarray}
{\cal L}^{soft}_{scalar}&=&-
m^{2}_{ \eta}(\eta^{ \dagger}\eta)-
m^{2}_{ \rho}(\rho^{ \dagger}\rho)-
m^{2}_{ \chi}(\chi^{ \dagger}\chi)-
m^{2}_{S}Tr[(S^{ \dagger}S)]
-m^{2}_{\eta^{\prime}}
(\eta^{\prime \dagger}\eta^{\prime})-
m^{2}_{\rho^{\prime}}
(\rho^{\prime \dagger}\rho^{\prime})
\nonumber \\ &-&
m^{2}_{\chi^{\prime}}
(\chi^{\prime \dagger}\chi^{\prime}) 
-m^{2}_{S^{\prime}}
Tr[(S^{\prime \dagger} S^{\prime})]
+[k_{1}(\epsilon \rho 
\chi \eta)+
k_{2}(\chi S \rho)
+
k^{\prime}_{1}(\epsilon 
\rho^{\prime}\chi^{\prime} 
\eta^{\prime})
\nonumber \\ 
&+&
k^{\prime}_{2}
(\chi^{\prime}S^{\prime} 
\rho^{\prime})+hc], 
\end{eqnarray}
$k_{1,2},k^{\prime}_{1}$ and $k^{\prime}_{2}$ has mass dimension.

The pattern of the symmetry breaking of the model is given by the following 
scheme
\begin{eqnarray}
&\mbox{MSUSY331}&
\stackrel{{\cal L}_{soft}}{\longmapsto}
\mbox{SU(3)}_C\ \otimes \ \mbox{SU(3)}_{L}\otimes \mbox{U(1)}_{N}
\stackrel{\langle\chi\rangle \langle \chi^{\prime}\rangle}{\longmapsto}
\mbox{SU(3)}_{C} \ \otimes \ \mbox{SU(2)}_{L}\otimes
\mbox{U(1)}_{Y} \nonumber \\
&\stackrel{\langle \rho, \eta, S \rho^{\prime},\eta^{\prime},S^{\prime}\rangle}{\longmapsto}&
\mbox{SU(3)}_{C} \ \otimes \ \mbox{U(1)}_{Q}.
\label{breaksusy331tou1}
\end{eqnarray}

\section{The scalar potential at MSUSY331}
\label{sec:sp}

The scalar potential is written as
\begin{equation}
V_{331}=V_D+V_F+V_{\mbox{soft}}
\label{ep1}
\end{equation}
where
\begin{eqnarray}
V_{D}&=&-{\cal L}_{D}, \nonumber \\
V_{F}&=&-{\cal L}_{F}=
\sum_{m}F^{\dagger}_{m} F_{m}, \nonumber \\ 
V_{soft}&=&-
{\cal L}^{\mbox{scalar}}_{soft}.
\label{ess}
\end{eqnarray}
How to get an expression to compare our scalar potential with 
the scalar potential in m331 model see App.(\ref{constpot}).

We will use below the following set of parameters in the scalar potential:
\begin{eqnarray}
f_{1}=1, \quad f_{3}=1.07,\quad f^{\prime}_{1}=f^{\prime}_{3}=10^{-6},\quad{\rm (dimensionless)}
\label{fs}
\end{eqnarray}
and
\begin{eqnarray}
-k_{1}=k^{\prime}_{1}=10,\;k_{3}=k^{\prime}_{3}=-100,\;  
-\mu_{\eta}=\mu_{\rho}=- \mu_{s}= \mu_{\chi}=1000, \quad \mbox{(in GeV)},
\label{ks}
\end{eqnarray}
we also use the constraint 
\begin{equation}
V^{2}_{\eta}+
V^{2}_{\rho}+2V^{2}_{2}=
(246\;{\rm GeV})^{2}
\end{equation} 
coming from $M_{W}$, where,
we have defined $V^{2}_{\eta}=v^{2}_{\eta}+
v^{\prime 2}_{\eta}$ and $V^{2}_{\rho}=
v^{2}_{\rho}+v^{\prime 2}_{\rho}$ and $V^{2}_{2}=v^{2}_{\sigma_2}+
v^{2}_{\sigma^{\prime}_{2}}$.
Assuming that $v_{\eta}=20$, 
$v_{\chi}=1500$, $v_{\sigma_{2}}=10$, 
$v^{\prime}_{\eta}=v^{\prime}_{\rho}=
v^{\prime}_{\sigma_{2}}=
v^{\prime}_{\chi}=1$ in GeV, the
value of $v_\rho$ is fixed by the constraint above. 

With this set of values for the parameters
the real mass eigenstates $H_{i}$ are 
(in GeV)
$M_{H^{0}_{i}}=125.5, 358.3, 506.3, 1207.4, 1496.5, 1818.1,4811.5,5531.4$, 
$i=1,\cdots,8$ and $M_{H_{j}}>M_{H_{i}}$ with $j>i$. 

In the pseudoscalar sector we have 
verified analytically that the mass 
matrix, has two
Goldstone bosons as it should. The other six physical pseudoscalars have the
following masses, with the same parameters as before, in GeV, 
$M_{A_{i}}=357.7,506.4,1207.4,1,496.5,4815.43,5531.5$ in GeV. The 
current $95\%$ CL mass bound on the lightest pseudoscalar at MSSM is 
$90.$GeV \cite{pdg}. 

In the Single charged Scalars we have two 
Goldstone bosons and nine massive states, 
in GeV. We get two unrelated bases. The first base is defined as
\begin{eqnarray}
\left(
\eta^{-}_{1},\rho^{-},\eta^{\prime -}_{1},\rho^{\prime -},h^{-}_{1},
h^{\prime -}_{1}
\right)^{T},
\label{base1single}
\end{eqnarray} 
we get the following masses 
$M_{H^{\pm}_{i}}=364.3,1206.3,1497.5,4811.7,5531.2$ in GeV. 
The second base is defined as
\begin{eqnarray}
\left(
\eta^{-}_{2},\chi^{-},\eta^{\prime -}_{2},\chi^{\prime -},h^{-}_{2},
h^{\prime -}_{2}
\right)^{T},
\label{base2single}
\end{eqnarray}
we get the following masses
$M_{H^{\pm}_{i}}=512.6,703.7,1132.9,4823.8,5509.7$ in GeV. The current 
$95\%$ CL mass bound on the lightest charged scalar at MSSM is 
$79.3$GeV \cite{pdg}.

In the Double charged Scalars we have one 
Goldstone bosons and seven massive states, 
in GeV, 
$M_{H^{\pm \pm}_{i}}=339.1,699.9,1134.3,
1256.3,1433.3,4799.3,4823.7$ in GeV. The current 
$95\%$ CL mass bound on the lightest doubly-charged scalar is 
$95$GeV and $100$GeV were obtained for left-right symmetric models (the exact 
limits depend on the leptons flavors) \cite{pdg}.

\begin{figure}[ht]
\begin{center}
\vglue -0.009cm
\mbox{\epsfig{file=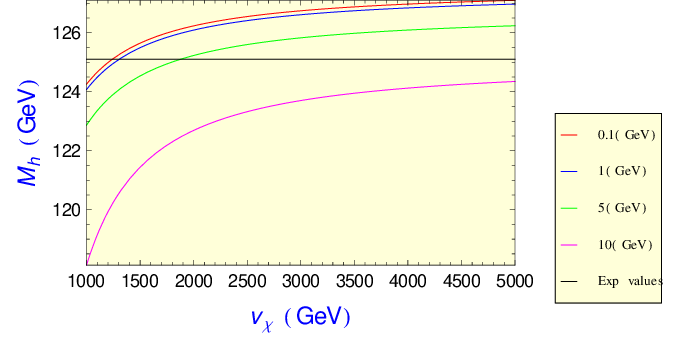,width=0.7\textwidth,angle=0}}       
\end{center}
\caption{The masses of light scalars, $M_{H^{0}_{1}}=M_{h}$, in terms of $v_{\chi}$ for 
some values of $v_{\chi^{\prime}}$ shown at box and Exp value is defined at 
Eq.(\ref{higgsexpvalue}). We are also using the 
parameters defined at Eqs.(\ref{fs},\ref{ks}).}
\label{fig5}
\end{figure}

\begin{figure}[ht]
\begin{center}
\vglue -0.009cm
\mbox{\epsfig{file=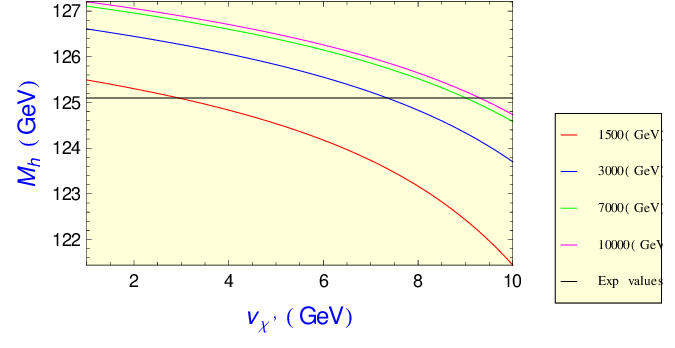,width=0.7\textwidth,angle=0}}       
\end{center}
\caption{The masses of light scalars, $M_{H^{0}_{1}}=M_{h}$, in terms of $v_{\chi^{\prime}}$ for 
some values of $v_{\chi}$ shown at box and Exp value is defined at 
Eq.(\ref{higgsexpvalue}). We are also using the 
parameters defined at Eqs.(\ref{fs},\ref{ks}).}
\label{fig6}
\end{figure}

\begin{figure}[ht]
\begin{center}
\vglue -0.009cm
\mbox{\epsfig{file=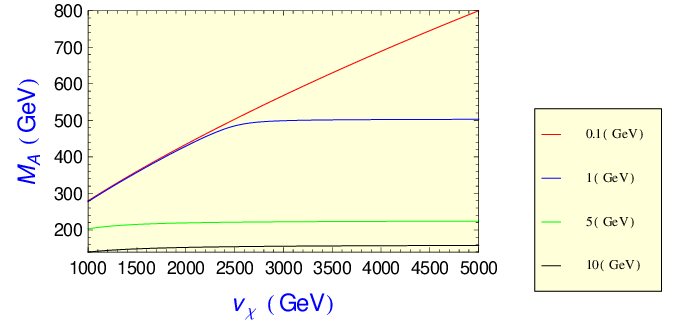,width=0.7\textwidth,angle=0}}       
\end{center}
\caption{The masses of light pseudoscalar, $M_{A_{1}}=M_{A}$, in terms of $v_{\chi}$ for 
some values of $v_{\chi^{\prime}}$ shown at box. We are also using the 
parameters defined at Eqs.(\ref{fs},\ref{ks}).}
\label{fig7}
\end{figure}

\begin{figure}[ht]
\begin{center}
\vglue -0.009cm
\mbox{\epsfig{file=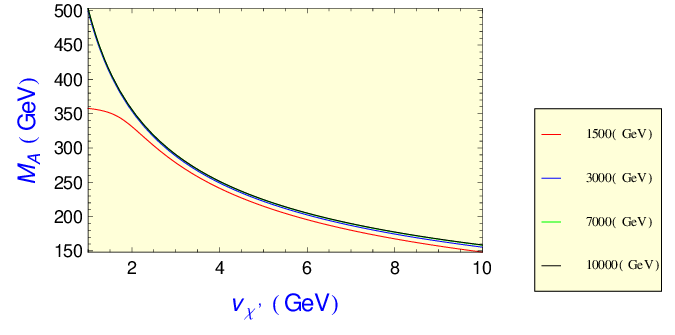,width=0.7\textwidth,angle=0}}       
\end{center}
\caption{The masses of light scalars, $M_{A_{1}}=M_{h}$, in terms of $v_{\chi^{\prime}}$ for 
some values of $v_{\chi}$ shown at box and Exp value is defined at 
Eq.(\ref{higgsexpvalue}). We are also using the 
parameters defined at Eqs.(\ref{fs},\ref{ks}).}
\label{fig8}
\end{figure}

\begin{figure}[ht]
\begin{center}
\vglue -0.009cm
\mbox{\epsfig{file=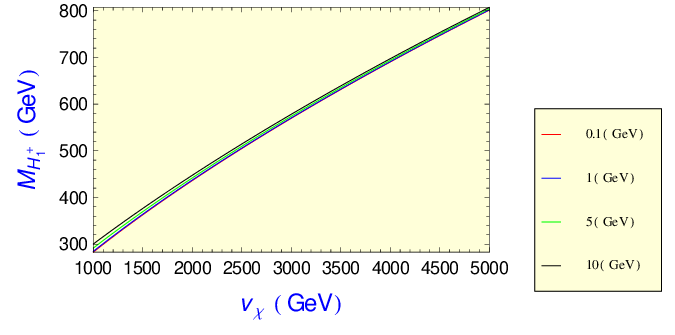,width=0.7\textwidth,angle=0}}       
\end{center}
\caption{The masses of single scalars at first base, 
$M_{H^{\pm}_{1}}$, see Eq.(\ref{base1single}), in terms of $v_{\chi}$ for 
some values of $v_{\chi^{\prime}}$ shown at box. We are also using the 
parameters defined at Eqs.(\ref{fs},\ref{ks}).}
\label{fig9}
\end{figure}

\begin{figure}[ht]
\begin{center}
\vglue -0.009cm
\mbox{\epsfig{file=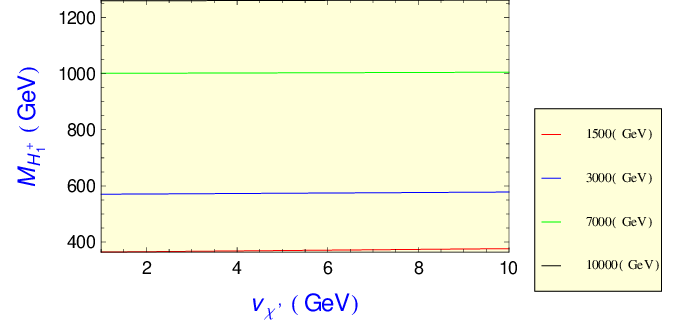,width=0.7\textwidth,angle=0}}       
\end{center}
\caption{The masses of single scalars at second base, $M_{H^{\pm}_{1}}$, see 
Eq.(\ref{base1single}), in terms of $v_{\chi^{\prime}}$ for 
some values of $v_{\chi}$ shown at box. We are also using the 
parameters defined at Eqs.(\ref{fs},\ref{ks}).}
\label{fig10}
\end{figure}

\begin{figure}[ht]
\begin{center}
\vglue -0.009cm
\mbox{\epsfig{file=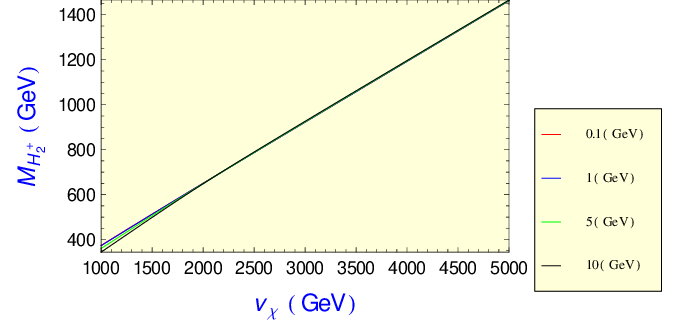,width=0.7\textwidth,angle=0}}       
\end{center}
\caption{The masses of single scalars at second base, $M_{H^{\pm}_{2}}$, see Eq.(\ref{base2single}), in terms of $v_{\chi}$ for 
some values of $v_{\chi^{\prime}}$ shown at box. We are also using the 
parameters defined at Eqs.(\ref{fs},\ref{ks}).}
\label{fig11}
\end{figure}

\begin{figure}[ht]
\begin{center}
\vglue -0.009cm
\mbox{\epsfig{file=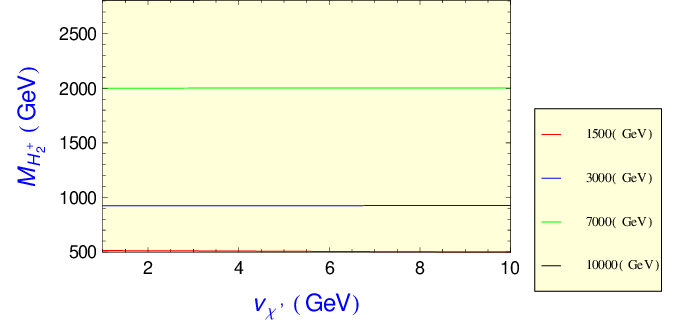,width=0.7\textwidth,angle=0}}       
\end{center}
\caption{The masses of single scalars at second base, $M_{H^{\pm}_{2}}$, see 
Eq.(\ref{base2single}), in terms of $v_{\chi^{\prime}}$ for 
some values of $v_{\chi}$ shown at box. We are also using the 
parameters defined at Eqs.(\ref{fs},\ref{ks}).}
\label{fig12}
\end{figure}

\begin{figure}[ht]
\begin{center}
\vglue -0.009cm
\mbox{\epsfig{file=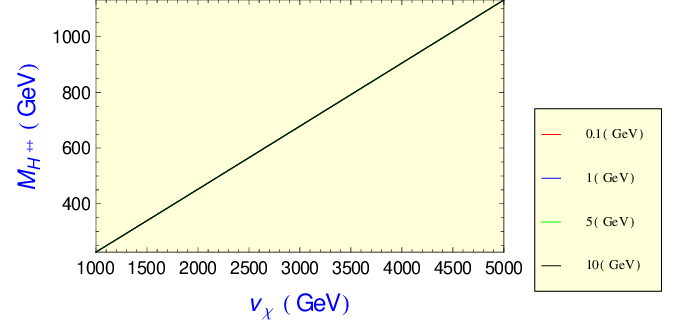,width=0.7\textwidth,angle=0}}       
\end{center}
\caption{The masses of double scalars, $M_{H^{\pm \pm}}$ in terms of 
$v_{\chi}$ for some values of $v_{\chi^{\prime}}$ shown at box. We are 
also using the parameters defined at Eqs.(\ref{fs},\ref{ks}).}
\label{fig13}
\end{figure}

\begin{figure}[ht]
\begin{center}
\vglue -0.009cm
\mbox{\epsfig{file=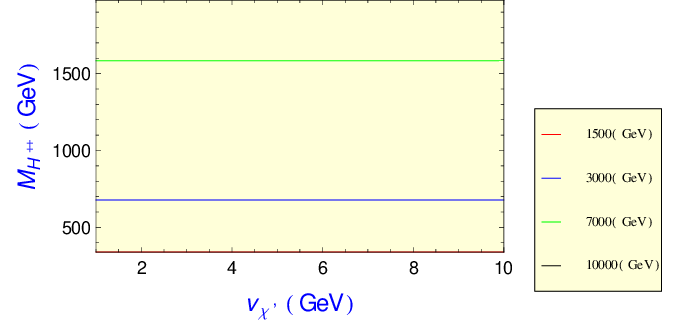,width=0.7\textwidth,angle=0}}       
\end{center}
\caption{The masses of double scalars, $M_{H^{\pm \pm}}$ in terms of $v_{\chi^{\prime}}$ for 
some values of $v_{\chi}$ shown at box. We are also using the 
parameters defined at Eqs.(\ref{fs},\ref{ks}).}
\label{fig14}
\end{figure}

We present next some plots to light scalar, first we fix 
some values for $v_{\chi^{\prime}}$ and present its 
mass in term of $v_{\chi}$, 
see Fig.(\ref{fig5}) and the same for pseudo-scalar see 
Fig.(\ref{fig7}), single charged higgses at first base, see Eq.(\ref{base1single}),  
Fig.(\ref{fig9}), second base, see Eq.(\ref{base2single}), at Fig.(\ref{fig11}) and double charged Higgs Fig.(\ref{fig13}) Next we fixed some values for $v_{\chi}$ and it is shown at 
Fig.(\ref{fig6}) and for pseudo-scalar at Fig.(\ref{fig8}), single charged higgses at first base, see Eq.(\ref{base1single}), 
Fig.(\ref{fig10}), second base, see Eq.(\ref{base2single}), at Fig.(\ref{fig12}) and double charged Higgs Fig.(\ref{fig14}).

\section{Conclusions}
\label{sec:conclusion}
We have studied the scalar potential of supersymmetric 
3-3-1 model with the sextet and anti-sextet and our results are in 
agreement with the acutal experimental data on scalar sector. Here we consider 
only the anti-sextet $S$ get VEV. We want to extend this analysis by 
also allowing the sextet $S^{\prime}$ to acquire vev, in similar way as 
we have done recently \cite{Rodriguez:2022hsj}. 

\begin{center}
{\bf Acknowledgments} 
\end{center}
We would like thanks V. Pleitez, J. C. Montero and 
B. L. Sch\'anchez-Vega for useful discussions above 3-3-1 models. We 
also to thanks IFT for the nice hospitality during my several visit 
to perform my studies about the severals 3-3-1 Models and also for done 
this article.
\appendix

\section{Review Scalar Potential at m331}
\label{psusy1}

Because of the anti-sextet, the scalar potential 
in the m331 become more complicated. The most general scalar potential involving 
triplets and the sextet is \cite{DeConto:2015eia}
\begin{eqnarray}
V(\eta,\rho,\chi,S)&=&V^{(2)}+V^{(3)}+ 
V^{(4a)}+V^{(4b)}+V^{(4c)}+V^{(4d)}+
V^{(4e)},
\end{eqnarray}
where
\begin{eqnarray}
V^{(2)}&=&
\mu_{1}^{2}(\eta^{\dagger}\eta) +
\mu_{2}^{2}(\rho^{\dagger}\rho) +
\mu_{3}^{2}(\chi^{\dagger}\chi) +
\mu_{4}^{2}Tr[(S^{\dagger} S)], \nonumber \\
V^{(3)}&=&
\frac{f_{1}}{3!}\varepsilon_{ijk}
\eta_{i}\rho_{j}\chi_{k}+
f_{2} \rho_{i}\chi_{j}S^{\dagger ij}+ 
f_{3}\eta_{i}\eta_{j}S^{\dagger ij}+
\frac{f_{4}}{3!}
\epsilon_{ijk}\epsilon_{lmn}
S_{il}S_{jm}S_{kn}+H.c., \nonumber \\
V^{(4a)}&=&
a_{1}(\eta^{\dagger}\eta)^{2} +
a_{2}(\rho^{\dagger}\rho)^{2}+
a_{3}(\chi^{\dagger}\chi)^{2}+
(\chi^{\dagger}\chi)\left[
a_{4}(\eta^{\dagger}\eta) +
a_{5}(\rho^{\dagger}\rho ) \right] +
a_{6}(\eta^{\dagger}\eta)(\rho^{\dagger}\rho) \nonumber \\ 
&+&
a_{7}(\chi^{\dagger}\eta)
(\eta^{\dagger}\chi) +
a_{8}(\chi^{\dagger}\rho)
(\rho^{\dagger}\chi) +
a_{9}(\eta^{\dagger}\rho)
(\rho^{\dagger}\eta) +
\left[
a_{10}(\chi^{\dagger}\eta)
(\rho^{\dagger}\eta)+H.c.
\right], \nonumber \\
V^{(4b)}&=&
b_{1}(\chi^{\dagger}S)(\hat{\chi}\eta) +
b_{2}(\rho^{\dagger}S)(\hat{\rho}\eta) +
b_{3}(\eta^{\dagger}S) \left[
(\hat{\chi}\rho) - (\hat{\rho}\chi)
\right]+H.c. , \nonumber \\
V^{(4c)}&=&
c_{1}Tr[ (\hat{\eta}S) (\hat{\eta}S)] +
c_{2}Tr[ (\hat{\rho}S) (\hat{\rho}S)] + 
H.c. , \nonumber \\
V^{(4d)}&=&
d_{1}(\chi^{\dagger}\chi)Tr[(S^{\dagger}S)] +
d_{2} Tr[(\chi^{\dagger}S)(S^{\dagger}\chi)] +
d_{3}(\eta^{\dagger}\eta)Tr[(S^{\dagger}S)] +
d_{4} Tr[(\eta^{\dagger}S)(S^{\dagger}\eta)] 
\nonumber \\
&+&
d_{5}(\rho^{\dagger}\rho)Tr[(S^{\dagger}S)] +
d_{6} Tr[(\rho^{\dagger}S)(S^{\dagger}\rho)], \nonumber \\
V^{(4e)}&=&
e_{1}(Tr[(S^{\dagger}S)])^{2} +
e_{2}Tr[(S^{\dagger}S)(S^{\dagger}S)],
\label{potentialm331}
\end{eqnarray}
and we have defined the $V^{(4b)}$ and 
$V^{(4c)}$ terms
\begin{eqnarray}
\hat{x}_{ij}\equiv \epsilon_{ijk}x_{k}, 
\end{eqnarray}
with $x= \eta, \rho, \chi$. The constants 
$f_{i}$, $i=1,2,3$ and $4$ have dimension of mass.

\section{Construction Scalar Potential}
\label{constpot}

To get the scalar potential of our model we have to eliminate the auxiliarly 
fields $F$ and $D$ that appear in our model. We are going to pick up the $F$ 
and $D$- terms we get
\begin{eqnarray}
{\cal L}^{Gauge}_{D}&=&\frac{1}{2}D^{a}D^{a}+ \frac{1}{2}DD \,\ , \nonumber \\
{\cal L}^{Scalar}_{F}&=& \vert F_{\eta} \vert^2+ \vert F_{\rho} \vert^2+ 
\vert F_{\chi} \vert^2+ \vert F_{S} \vert^2 + 
\vert F_{\eta^{\prime}} \vert^2+ 
\vert F_{\rho^{\prime}} \vert^2+ 
\vert F_{\chi^{\prime}} \vert^2+ \vert F_{S^{\prime}} \vert^2, \nonumber \\
{\cal L}^{Scalar}_{D}&=& \frac{g}{2} \left[ \bar{\eta}\lambda^a\eta+ 
\bar{\rho}\lambda^a\rho+ \bar{\chi}\lambda^a\chi+ 
\bar{S}\lambda^aS- 
\bar{\eta}^{\prime}\lambda^{* a}\eta^{\prime}- 
\bar{\rho}^{\prime}\lambda^{* a}\rho^{\prime}- 
\bar{\chi}^{\prime}\lambda^{* a}\chi^{\prime}- 
\bar{S}^{\prime}\lambda^{* a}S^{\prime} \right] D^{a} \nonumber \\
&+& \frac{g^{ \prime}}{2} \left[  
\bar{\rho}\rho- \bar{\chi}\chi- \bar{\rho}^{\prime}\rho^{\prime}+ 
\bar{\chi}^{\prime}\chi^{\prime} \right]D, \nonumber \\
{\cal L}^{W2}_{F}&=& \frac{1}{2} \left[ \mu_{ \eta} ( \eta F_{\eta^{\prime}}+ 
\eta^{\prime} F_{ \eta}+ \bar{\eta} \bar{F}_{\eta^{\prime}}+ 
\bar{\eta^{\prime}} \bar{F}_{ \eta})+ 
\mu_{ \rho} ( \rho F_{\rho^{\prime}}+ \rho^{\prime} F_{ \rho}+
\bar{\rho} \bar{F}_{\rho^{\prime}}+ \bar{\rho^{\prime}} \bar{F}_{ \rho})  
\right. \nonumber \\
&+& \left. \mu_{ \chi} ( \chi F_{\chi^{\prime}}+ 
\chi^{\prime} F_{ \chi}+
\bar{\chi} \bar{F}_{\chi^{\prime}}+ \bar{\chi^{\prime}} \bar{F}_{ \chi}) 
\right], \nonumber \\
{\cal L}^{W3}_{F}&=& \vert F_{\eta} \vert^2+ \vert F_{\rho} \vert^2+ 
\vert F_{\chi} \vert^2+ \vert F_{S} \vert^2 + 
\vert F_{\eta^{\prime}} \vert^2+ 
\vert F_{\rho^{\prime}} \vert^2+ 
\vert F_{\chi^{\prime}} \vert^2+ \vert F_{S^{\prime}} \vert^2. 
\nonumber \\
\end{eqnarray}

From the equation described above we can construct
\begin{eqnarray}
{\cal L}_{F}&=&{\cal L}^{Scalar}_{F}+
{\cal L}^{W2}_{F}+{\cal L}^{W3}_{F} 
\nonumber \\
&=&\vert F_{\eta} \vert^2+ \vert F_{\rho} \vert^2+ 
\vert F_{\chi} \vert^2+  
\vert F_{\eta^{\prime}} \vert^2+ 
\vert F_{\rho^{\prime}} \vert^2+ 
\vert F_{\chi^{\prime}} \vert^2 \nonumber \\
&+& \frac{1}{2} \left[ \mu_{ \eta} ( \eta F_{\eta^{\prime}}+ 
\eta^{\prime} F_{ \eta}+ \bar{\eta} \bar{F}_{\eta^{\prime}}+ 
\bar{\eta^{\prime}} \bar{F}_{ \eta})+ 
\mu_{ \rho} ( \rho F_{\rho^{\prime}}+ \rho^{\prime} F_{ \rho}+
\bar{\rho} \bar{F}_{\rho^{\prime}}+ \bar{\rho^{\prime}} \bar{F}_{ \rho})  
\right. \nonumber \\
&+& \left. \mu_{ \chi} ( \chi F_{\chi^{\prime}}+ 
\chi^{\prime} F_{ \chi}+
\bar{\chi} \bar{F}_{\chi^{\prime}}+ \bar{\chi^{\prime}} \bar{F}_{ \chi}) 
\right] + \frac{1}{3} \left[ f_{1} \epsilon (F_{ \rho} \chi \eta+ \rho F_{ \chi} \eta+ \rho \chi F_{ \eta} \right. \nonumber \\
&+& \left.
\bar{F}_{ \rho} \bar{\chi} \bar{\eta}+ \bar{\rho} \bar{F}_{ \chi} \bar{\eta}+ 
\bar{\rho} \bar{\chi} \bar{F}_{ \eta} ) + f^{\prime}_{1} \epsilon (
F_{ \rho^{\prime}} \chi^{\prime} \eta^{\prime}+ 
\rho^{\prime} F_{ \chi^{\prime}} \eta^{\prime}+ 
\rho^{\prime} \chi^{\prime} F_{ \eta^{\prime}} \right. \nonumber \\
&+& \left.
\bar{F}_{ \rho^{\prime}} \bar{\chi^{\prime}} \bar{\eta^{\prime}}+ 
\bar{\rho^{\prime}} \bar{F}_{ \chi^{\prime}} \bar{\eta^{\prime}}+ 
\bar{\rho^{\prime}} \bar{\chi^{\prime}} \bar{F}_{ \eta^{\prime}})
 \right] \,\ , \nonumber \\
{\cal L}_{D}&=&{\cal L}^{Gauge}_{D}+
{\cal L}^{Scalar}_{D} \nonumber \\
&=& \frac{1}{2}D^{a}D^{a}+ \frac{1}{2}DD 
+ \frac{g}{2} \left[ \bar{\eta}\lambda^a\eta+ 
\bar{\rho}\lambda^a\rho+ \bar{\chi}\lambda^a\chi- 
\bar{\eta}^{\prime}\lambda^{* a}\eta^{\prime}- 
\bar{\rho}^{\prime}\lambda^{* a}\rho^{\prime} \right. \nonumber \\
&-& \left. 
\bar{\chi}^{\prime}\lambda^{* a}\chi^{\prime} \right] D^{a}+ 
\frac{g^{ \prime}}{2} \left[  
\bar{\rho}\rho- \bar{\chi}\chi- \bar{\rho}^{\prime}\rho^{\prime}+ 
\bar{\chi}^{\prime}\chi^{\prime} \right]D.
\label{auxiliarm1}
\end{eqnarray}

We will now show that these fields can be eliminated through the 
Euler-Lagrange equations
\begin{eqnarray}
\frac{\partial {\cal L}}{\partial \phi}- \partial_{m} 
\frac{\partial {\cal L}}{\partial (\partial_{m} \phi)}=0 \,\ ,  
\label{Euler-Lagrange Equation}
\end{eqnarray}
where 
$\phi = \eta , \rho , \chi , \eta^{\prime}, \rho^{\prime}, \chi^{\prime}$. 
Formally auxiliary fields are defined 
as fields having no kintetic terms. Thus, this definition immediately yields 
that the Euler-Lagrange equations for auxiliary fields simplify to 
$\frac{\partial {\cal L}}{\partial \phi}=0$.

\begin{eqnarray}
V_D&=&-{\cal L}_D=\frac{1}{2}\left(D^aD^a+DD\right)\nonumber \\ &=&
\frac{g^{\prime2}}{2}(\rho^\dagger\rho-\rho^{\prime\dagger}\rho^\prime
-\chi^\dagger\chi+\chi^{\prime\dagger}\chi^\prime)^2+
\frac{g^2}{8}\sum_{i,j}\left(\eta^\dagger_i\lambda^a_{ij}\eta_j
+\rho^\dagger_i\lambda^a_{ij}\rho_j
+\chi^\dagger_i\lambda^a_{ij}\chi_j+S^\dagger_{ij}\lambda^a_{jk}
S_{kl}\right.
\nonumber \\ &-&
\left.\eta^{\prime\dagger}_i\lambda^{*a}_{ij}\eta^\prime_j 
-\rho^{\prime\dagger}_i\lambda^{*a}_{ij}\rho^\prime_j
-\chi^{\prime\dagger}_i\lambda^{*a}_{ij}\chi^\prime_j-
S^{\prime\dagger}_{ij}\lambda^{*a}_{jk} S^\prime_{kl} \right)^2,
\label{esd}
\end{eqnarray}
We can use the following relations
\begin{eqnarray}
\lambda^{a}_{ij}\lambda^{a}_{kl}= 
\frac{(-2)}{3}\delta_{ij}\delta_{kl}+
2\delta{il}\delta_{jk}.
\end{eqnarray}
Then
\begin{eqnarray}
\eta^{\dagger}_{i}\eta_{j}
\eta^{\dagger}_{k}\eta_{l}
\lambda^{a}_{ij}\lambda^{a}_{kl}=- 
\frac{2}{3}
\left( \eta^{\dagger}\eta \right)
\left( \eta^{\dagger}\eta \right)+ 2 
\left( \eta^{\dagger}\eta \right)
\left( \eta^{\dagger}\eta \right)= 
\frac{4}{3} \left( \eta^{\dagger}\eta \right)^{2},
\end{eqnarray}
in similar way
\begin{eqnarray}
\rho^{\dagger}_{i}\rho_{j}
\rho^{\dagger}_{k}\rho_{l}
\lambda^{a}_{ij}\lambda^{a}_{kl}&=& 
\frac{4}{3} \left( \rho^{\dagger}\rho \right)^{2}, \nonumber \\
\chi^{\dagger}_{i}\chi_{j}
\chi^{\dagger}_{k}\chi_{l}
\lambda^{a}_{ij}\lambda^{a}_{kl}&=& 
\frac{4}{3} \left( \chi^{\dagger}\chi \right)^{2}, \nonumber \\
S^{\dagger}_{il}S_{mj}
S^{\dagger}_{kn}S_{op}
\lambda^{a}_{lm}\lambda^{a}_{no}&=&- 
\frac{2}{3}
{\mbox Tr}\left[ 
\left( S^{\dagger}S \right)
\left( S^{\dagger}S \right) \right] + 2 
{\mbox Tr} \left[ (S^{\dagger}S) 
(S^{\dagger}S) \right].
\end{eqnarray}

By another hand
\begin{eqnarray}
\eta^{\dagger}_{i}\eta_{j}
\rho^{\dagger}_{k}\rho_{l}
\lambda^{a}_{ij}\lambda^{a}_{kl}&=&- 
\frac{2}{3}
\left( \eta^{\dagger}\eta \right)
\left( \rho^{\dagger}\rho \right)+ 2 
\left( \rho^{\dagger}\eta \right)
\left( \eta^{\dagger}\rho \right), 
\nonumber \\
\eta^{\dagger}_{i}\eta_{j}
\chi^{\dagger}_{k}\chi_{l}
\lambda^{a}_{ij}\lambda^{a}_{kl}&=&- 
\frac{2}{3}
\left( \eta^{\dagger}\eta \right)
\left( \chi^{\dagger}\chi \right)+ 2 
\left( \chi^{\dagger}\eta \right)
\left( \eta^{\dagger}\chi \right), 
\nonumber \\
\rho^{\dagger}_{i}\rho_{j}
\chi^{\dagger}_{k}\chi_{l}
\lambda^{a}_{ij}\lambda^{a}_{kl}&=&- 
\frac{2}{3}
\left( \rho^{\dagger}\rho \right)
\left( \chi^{\dagger}\chi \right)+ 2 
\left( \chi^{\dagger}\rho \right)
\left( \rho^{\dagger}\chi \right),
\end{eqnarray}

We can rewrite $V_{D}$ in the following way
\begin{eqnarray}
V_{D}&=&\frac{g^{\prime2}}{2}
\left[ (\rho^{\dagger}\rho)^{2}+
(\chi^{\dagger}\chi)^{2}-
2(\rho^{\dagger}\rho)(\chi^{\dagger}\chi)
+ \ldots  \right] \nonumber \\
&+& \frac{g^{2}}{8}\left\{
\frac{(-4)}{3}\left[
(\eta^{\dagger}\eta)^{2}+
(\rho^{\dagger}\rho)^{2}+
(\chi^{\dagger}\chi)^{2}+
(\eta^{\dagger}\eta)(\rho^{\dagger}\rho)+
(\eta^{\dagger}\eta)(\chi^{\dagger}\chi)
\right. \right. \nonumber \\ 
&+& \left. \left.
(\eta^{\dagger}\eta)Tr[(S^{\dagger}S)]+
(\rho^{\dagger}\rho)(\chi^{\dagger}\chi)
+
(\rho^{\dagger}\rho)Tr[(S^{\dagger}S)]+ 
(\chi^{\dagger}\chi)Tr[(S^{\dagger}S)]
\right]
\right. \nonumber \\
&+& \left. 
4 \left[
(\eta^{\dagger}\rho)(\rho^{\dagger}\eta)+
(\eta^{\dagger}\chi)(\chi^{\dagger}\eta)
+
Tr[(S^{\dagger}\eta)(\eta^{\dagger}S)]+
(\rho^{\dagger}\chi)(\chi^{\dagger}\rho)
\right. \right. \nonumber \\
&+& \left. \left.
Tr[(S^{\dagger}\rho)(\rho^{\dagger}S)]+
Tr[(S^{\dagger}\chi)(\chi^{\dagger}S)]
\right] 
- \frac{2}{3}
\left(Tr[(S^{\dagger}S)]\right)^{2}
\right.  \nonumber \\
&+& \left.  2 
Tr[(S^{\dagger}S)(S^{\dagger}S)]
+ \ldots
\right\}.
\end{eqnarray}
where $\ldots$ include the new fields 
due the SUSY algebra. As we want to 
compare our potential with the m331 
we omitt those terms.

\begin{eqnarray}
V_F&=&-{\cal L}_F=\sum_mF^*_m F_m\nonumber \\ &=&
\sum_{i,j,k}\left[\left\vert \frac{\mu_\eta}{2} \eta^{\prime}_i+\frac{f_1}{3}
\epsilon_{ijk}\rho_j\chi_k
+\frac{2f_2}{3}\eta_iS_{ij}\right\vert^2+
\left\vert \frac{\mu_\rho}{2}\rho^\prime_i+\frac{f_1}{3}
\epsilon_{ijk}\chi_j\eta_k+\frac{f_3}{3}\chi_i S_{ij}\right\vert^2
\right.\nonumber \\ &+&\left.
\left\vert
\frac{\mu_\chi}{2}\chi^\prime_i+\frac{f_1}{3}\epsilon_{ijk}\rho_j\eta_k 
+\frac{f_3}{3}\rho_i S_{ij}\right\vert^2
+\left\vert \frac{\mu_s}{2}S^\prime_{ij}+\frac{f_2}{3}\eta_i\eta_j+\frac{f_3}{3}
\chi_i\rho_j\right\vert^2\right.\nonumber \\ &+&\left.\left\vert 
\frac{\mu_\eta}{2} \eta_i+\frac{f^\prime_1}{3}
\epsilon_{ijk}\rho^\prime_j\chi^\prime_k
+\frac{2f^\prime_2}{3}\eta^\prime_iS^\prime_{ij}\right\vert^2+
\left\vert \frac{\mu_\rho}{2}\rho_i+\frac{f^\prime_1}{3}
\epsilon_{ijk}\chi^\prime_j\eta^\prime_k+\frac{f^\prime_3}{3}\chi^\prime_i
S^\prime_{ij} 
\right\vert^2\right.\nonumber \\ &+&\left.
\left\vert
\frac{\mu_\chi}{2}\chi_i+\frac{f^\prime_1}{3}\epsilon_{ijk}\rho^\prime_j
\eta^\prime_k
+\frac{f^\prime_3}{3}\rho^\prime_i S^\prime_{ij}\right\vert^2
+\left\vert
\frac{\mu_s}{2}S_{ij}+\frac{f^\prime_2}{3}\eta^\prime_i\eta^\prime_j+ 
\frac{f^\prime_3}{3}
\chi^\prime_i\rho^\prime_j\right\vert^2
\right] \nonumber \\
\end{eqnarray}
We can use the following relation
\begin{equation}
\epsilon_{ijk}\epsilon_{kmn}=
\delta_{im}\delta_{jn}-
\delta_{in}\delta_{jm}.
\end{equation}

In both case $f_{3}=0=f^{\prime}_{3}$, 
then we get
\begin{eqnarray}
\left\vert 
\frac{\mu_\eta}{2} \eta_i + \ldots 
\right\vert^2 
&=& \frac{\mu_{\eta}}{4}
(\eta^{\dagger}\eta), \,\
\left\vert \frac{f_1}{3}
\epsilon_{ijk}\rho_j\chi_k + 
\ldots \right\vert^2 =
\frac{f^{2}_{1}}{9}\left[
(\rho^{\dagger}\rho)(\chi^{\dagger}\chi)- 
(\rho^{\dagger}\chi)(\chi^{\dagger}\rho) 
\right], \nonumber \\
\left\vert 
\frac{\mu_\rho}{2} \rho_i + \ldots 
\right\vert^2 
&=& \frac{\mu^{2}_{\rho}}{4}
(\rho^{\dagger}\rho), \,\ 
\left\vert 
\frac{\mu_\chi}{2} \chi_i + \ldots 
\right\vert^2 
= \frac{\mu^{2}_{\chi}}{4}
(\chi^{\dagger}\chi), \nonumber \\
\left\vert \frac{f_1}{3}
\epsilon_{ijk}\chi_j\eta_k+\frac{f_3}{3}\chi_i S_{ij}+ \ldots \right\vert^2 
&=& \frac{f^{2}_{1}}{9} \left[ 
( \chi^{\dagger}\chi )
( \eta^{\dagger}\eta )-
( \chi^{\dagger}\eta )
( \eta^{\dagger}\chi ) 
\right]+
\frac{f_{1}f_{2}}{9} \left[ 
\epsilon ( \chi^{\dagger}\eta^{\dagger})
( \chi S)+H.c. 
\right] \nonumber \\ &+&
\frac{f^{2}_{2}}{9} {\mbox Tr}\left[ 
(S^{\dagger}\chi )( \chi^{\dagger}S) 
\right], \nonumber \\
\left\vert \frac{f_1}{3}
\epsilon_{ijk}\rho_j\eta_k+\frac{f_3}{3}\rho_i S_{ij}+ \ldots \right\vert^2 
&=& \frac{f^{2}_{1}}{9} \left[ 
( \rho^{\dagger}\rho )
( \eta^{\dagger}\eta )-
( \rho^{\dagger}\eta )
( \eta^{\dagger}\rho ) 
\right]+
\frac{f_{1}f_{2}}{9} \left[ 
\epsilon ( \rho^{\dagger}\eta^{\dagger})
( \rho S)+H.c. 
\right] \nonumber \\ &+&
\frac{f^{2}_{2}}{9} {\mbox Tr}\left[ 
(S^{\dagger}\rho )( \rho^{\dagger}S) 
\right], \nonumber \\
\left\vert
\frac{\mu_s}{2}S_{ij}+ \ldots 
\right\vert^2 &=& \frac{\mu^{2}_s}{4} 
(Tr[(S^{\dagger}S)])^{2}, \nonumber \\ 
\left\vert \frac{f_1}{3}
\chi_i\rho_j + \ldots \right\vert^2 
&=& \frac{f^{2}_1}{9} \left[
(\rho^{\dagger}\rho)(\chi^{\dagger}\chi)-
(\rho^{\dagger}\chi)(\chi^{\dagger}\rho)
\right]. \nonumber \\
\end{eqnarray}

Therefore
\begin{eqnarray}
V_F&=& 
\frac{\mu^{2}_{\eta}}{4}(\eta^{\dagger}\eta)^{2}+
\frac{\mu^{2}_{\rho}}{4}(\rho^{\dagger}\rho)^{2}+
\frac{\mu^{2}_{\chi}}{4}(\chi^{\dagger}\chi)^{2}+ 
\frac{\mu^{2}_{s}}{4}
(Tr[(S^{\dagger}S)])^{2}
\nonumber \\ &+&
\frac{f^{2}_{1}}{9}\left[
(\eta^{\dagger}\eta)(\rho^{\dagger}\rho)-
(\rho^{\dagger}\eta)(\eta^{\dagger}\rho)+
(\eta^{\dagger}\eta)(\chi^{\dagger}\chi)
\right. \nonumber \\
&-& \left.
(\chi^{\dagger}\eta)(\eta^{\dagger}\chi)+
(\rho^{\dagger}\rho)(\chi^{\dagger}\chi)-
(\rho^{\dagger}\chi)(\chi^{\dagger}\rho)
\right]+ 
\frac{f^{2}_{2}}{9}\left\{
Tr[(S^{\dagger}\rho)(\rho^{\dagger}S)]+
Tr[(S^{\dagger}\chi)(\chi^{\dagger}S)]
\right\}
\nonumber \\ &+&
\left\{ \frac{f_{1}f_{2}}{9}
\left[ 
\epsilon (\rho^{\dagger}\eta^{\dagger}) 
(\rho S)+
\epsilon (\chi^{\dagger}\eta^{\dagger}) 
(\chi S) \right]+H.c.\right\}
+ \ldots ,
\label{esf}
\end{eqnarray}

The scalar potential in this model is
\begin{eqnarray}
V_{scalar}&=&
m^{2}_{ \eta}(\eta^{ \dagger}\eta)+
m^{2}_{ \rho}(\rho^{ \dagger}\rho)+
m^{2}_{ \chi}(\chi^{ \dagger}\chi)+
m^{2}_{S}Tr[(S^{ \dagger}S)]
-[k_{1}\epsilon \rho \chi \eta +
k_{2}\chi \rho S^{\dagger}+ H.c.] \nonumber \\
&+&
\left( \frac{\mu^{2}_{\eta}}{4}- \frac{g^{2}}{6}
\right) (\eta^{\dagger}\eta)^{2}+
\left( \frac{\mu^{2}_{\rho}}{4}+ 
\frac{g^{\prime 2}}{2}- \frac{g^{2}}{6}
\right) (\rho^{\dagger}\rho)^{2}+
\left( \frac{\mu^{2}_{\chi}}{4}+ 
\frac{g^{\prime 2}}{2}- \frac{g^{2}}{6}
\right) (\chi^{\dagger}\chi)^{2}
\nonumber \\ &+&
\left(
\frac{f^{2}_{1}}{9}- \frac{g^{2}}{6}
\right)(\eta^{\dagger}\eta)(\rho^{\dagger}\rho)+
\left(
\frac{f^{2}_{1}}{9}- \frac{g^{2}}{6}
\right)(\eta^{\dagger}\eta)(\chi^{\dagger}\chi)+
\left(
\frac{f^{2}_{1}}{9}-g^{\prime 2}- \frac{g^{2}}{6}
\right)(\rho^{\dagger}\rho)(\chi^{\dagger}\chi)
\nonumber \\ &+&
\left(
\frac{g^{2}}{2}-\frac{f^{2}_{1}}{9}
\right)(\eta^{\dagger}\rho)(\rho^{\dagger}\eta)+
\left(
\frac{g^{2}}{2}-\frac{f^{2}_{1}}{9}
\right)(\eta^{\dagger}\chi)(\chi^{\dagger}\eta)+
\left(
\frac{g^{2}}{2}-\frac{f^{2}_{1}}{9}
\right)(\rho^{\dagger}\chi)(\chi^{\dagger}\rho) \nonumber \\
&-& \frac{g^{2}}{6} \left[
(\eta^{\dagger}\eta)Tr[(S^{\dagger}S)]+
(\rho^{\dagger}\rho)Tr[(S^{\dagger}S)]+ 
(\chi^{\dagger}\chi)Tr[(S^{\dagger}S)] 
\right] \nonumber \\
&+& 
\frac{g^{2}}{2}
Tr[(S^{\dagger}\eta)(\eta^{\dagger}S)]+
\left( \frac{g^{2}}{2}+
\frac{f^{2}_{2}}{9} \right)
\left\{
Tr[(S^{\dagger}\rho)(\rho^{\dagger}S)]+
Tr[(S^{\dagger}\chi)(\chi^{\dagger}S)]
\right\} \nonumber \\
&+&
\left( \frac{\mu^{2}_{s}}{4}-
\frac{g^{2}}{2} \right)
\left(Tr[(S^{\dagger}S)]\right)^{2}
+g^{2} Tr[(S^{\dagger}S)(S^{\dagger}S)] 
\nonumber \\
&+&
\left\{ \frac{f_{1}f_{2}}{9}
\left[ 
\epsilon (\rho^{\dagger}\eta^{\dagger}) 
(\rho S)+
\epsilon (\chi^{\dagger}\eta^{\dagger}) 
(\chi S) \right]+H.c.\right\}
+ \ldots
\end{eqnarray}
and comparing it with Eq.(\ref{potentialm331}) we can see
\begin{eqnarray}
\mu^{2}_{1}&=& m^{2}_{ \eta}, \quad 
\mu^{2}_{2}= m^{2}_{ \rho}, \quad
\mu^{2}_{3}= m^{2}_{ \chi}, \quad
\mu^{2}_{4}= m^{2}_{S}, \nonumber \\ 
f_{1}&=&-3!k_{1}, \quad
f_{2}=-3!k_{2},
\nonumber \\
a_{1}&=&\frac{\mu^{2}_{\eta}}{4}- \frac{g^{2}}{6}, 
\quad 
e_{1}=\frac{\mu^{2}_{s}}{4}-\frac{g^{2}}{2}, \quad
e_{2}=g^{2}, \nonumber \\
a_{2}&=&\frac{\mu^{2}_{\rho}}{4}+ \frac{g^{\prime 2}}{2}
- \frac{g^{2}}{6},
a_{3}=\frac{\mu^{2}_{\chi}}{4}+ \frac{g^{\prime 2}}{2}
- \frac{g^{2}}{6}, \nonumber \\
a_{4}&=&- \frac{g^{2}}{6}+
\frac{f^{2}_{1}}{9}=a_{6}=a_{7}=a_{8}=a_{9}, \nonumber \\
a_{5}&=&- g^{\prime 2}- \frac{g^{2}}{6}+
\frac{f^{2}_{1}}{9}, \nonumber \\
d_{1}&=&- \frac{g^{2}}{6}=d_{3}=d_{4}=d_{5}, \nonumber \\
d_{2}&=&d_{4}=d_{6}=\frac{g^{2}}{2}, 
\nonumber \\
a_{10}&=&=0.
\label{comppotnonsusycomsusy}
\end{eqnarray}

We get analytical two Goldstone bosons at CP-odd sector, two Goldstone 
bosons at Sngly charged sector and one Goldstone boson at Doubly 
charged sector. We will present our numerical rsults 
in Sec.(\ref{sec:sp}).



\begin{thebibliography}{99}


\bibitem{singer}M. Singer, J. W. F. Valle, and J. Schechter, 
{\it Canonical neutral-current predictions from the weak-electromagnetic gauge group $SU(3)\otimes U(1)$}, 
{\sl Phys. Rev.} {\bf D 22}, 738, (1980).
\bibitem{ppf}F. Pisano and V. Pleitez, 
{\it An $SU(3)\otimes U(1)$ model for electroweak interactions}, 
{\sl Phys. Rev.}  {\bf D 46}, 410, (1992); 
R. Foot, O. F. Hernandez, F. Pisano and V. Pleitez, 
{\it Lepton masses in an $SU(3)_{L} \otimes U(1)_{N}$ gauge model},
{\sl Phys. Rev.} {\bf D 47}, 4158, (1993).

\bibitem{331rh} R. Foot, H.N. Long and Tuan A. Tran,
{\it $SU(3)_{L}\otimes U(1)_{N}$ and $SU(4)_{L}\otimes U(1)_{N}$ gauge models with right-handed neutrinos}, 
{\sl Phys. Rev.} {\bf D 50}, 34, (1994);
J. C. Montero, F. Pisano and V. Pleitez, {\sl Phys. Rev.} {\bf D 47}, 2918, (1993); 
H.N. Long, 
{\it $SU(3)_{L}\otimes U(1)_{N}$ model for right-handed neutrino neutral currents}, 
{\sl Phys. Rev.} {\bf D 54}, 4691, (1996); 
H.N. Long,
{\it $SU(3)_{C}\otimes SU(3)_{L}\otimes U(1)_{N}$ model with right-handed neutrinos}, 
{\sl Phys. Rev.} {\bf D 53}, 437, (1996).

\bibitem{Pleitez:1994pu}V. Pleitez,
{\it New fermions and a vector - like third generation in $SU(3)_{C}\otimes SU(3)_{L}\otimes U(1)_{N}$ models},
{\sl Phys. Rev.}{\bf D53}, 514, (1996).
\bibitem{Ponce:2001jn}
W. A. Ponce, J. B. Florez and L. A. Sanchez,
{\it Analysis of SU(3)(c) x SU(3)(L) x U(1)(X) local gauge theory},
{\sl Int. J. Mod. Phys.} {\bf A17}, 643, (2002).

\bibitem{Sanchez:2001ua}L. A. Sanchez, W. A. Ponce and R. Martinez,
{\it SU(3) ($c$) x SU(3) ($\ell$) x U(1) ($X$) as an E(6) subgroup},
{\sl Phys. Rev.} {\bf D64}, 075013, (2001).

\bibitem{Martinez:2001mu}
R. Martinez, W. A. Ponce and L. A. Sanchez,
{\it SU(3) (C) x SU(3) (L) x U(1) ($X$) as an SU(6) x U(1) ($X$) subgroup},
{\sl Phys. Rev.} {\bf D65}, 055013, (2002).


\bibitem{sg}S. J. L. Rosner,
{\it Resource letter: The Standard model and beyond},
{\sl Am. J. Phys.} {\bf 71}, 302, (2003); 
A. S. Kronfeld and C. Quigg,
{\it Resource Letter: Quantum Chromodynamics},
{\sl Am. J. Phys.}{\bf 78}, 1081, (2010).

\bibitem{ema1} T. V. Duong and E. Ma, 
{\it Supersymmetric $SU(3) \otimes U(1)$ Gauge Model: 
Higgs Structure at the Electroweak Energy Scale}, 
{\sl Phys. Lett.}{\bf B316}, 307 (1993).
\bibitem{pal2} H. N. Long and P. B. Pal, {\it Nucleon instability in a supersymmetric 
$SU(3)_{C}\otimes SU(3)_{L}\otimes U(1)$ model}, 
{\sl Mod. Phys. Lett.}{\bf A13}, 2355, (1998).


\bibitem{331susy1}J. C. Montero, V. Pleitez and M. C. Rodriguez, 
{\it A Supersymmetric 3-3-1 model},
{\sl Phys. Rev.} {\bf D65}, 035006, (2002).
\bibitem{mcr} M. Capdequi-Peyran\`ere and M.C. Rodriguez, 
{\it Charginos and neutralinos production at 3-3-1 supersymmetric model in e- e- scattering},
{\sl Phys. Rev.} {\bf D 65}, 035001 (2002);
M. C. Rodriguez,
{\it Double Chargino Production in $e^{-}e^{-}$ scattering},
{\sl Int. J. Mod. Phys. }{\bf A 22}, 6080, (2008).
\bibitem{Rodriguez:2005jt}M. C. Rodriguez,
{\it Scalar sector in the minimal supersymmetric 3-3-1 model},
{\sl Int. J. Mod. Phys.}{\bf A21}, 4303, (2006).
\bibitem{Rodriguez:2010tn}M. C. Rodriguez,
\emph{Mass Spectrum in the Minimal Supersymmetric 3-3-1 model},
\emph{J. Mod. Phys.}{\bf 2}, 1193, (2011).
\bibitem{Rodriguez:2022hsj}M. C. Rodriguez,
{\it Gauge bosons masses in the context of the Supersymmetric $SU(3)_{C}\otimes SU(3)_{L}\otimes U(1)_{N}$ Model},
{\sl Int. J. Mod. Phys.}{\bf A}, 2440001, (2024);
[arXiv:2209.04653 [hep-ph]].

\bibitem{Pleitez:1992xh}V. Pleitez and M. D. Tonasse,
{\it Heavy charged leptons in an 
$SU(3)_{L}\otimes U(1)_{N}$ model},
{\sl Phys. Rev.} {\bf D 48}, 2353, (1993).
\bibitem{Pleitez:1993gc}V. Pleitez and M. D. Tonasse,
{\it Neutrinoless double beta decay in an SU(3)-L x U(1)-N model},
{\sl Phys. Rev.}{\bf D48}, 5274, (1993).

\bibitem{331susy2} J. C. Montero, V. Pleitez and M. C. Rodriguez, 
{\it Supersymmetric 3-3-1 model with right-handed
neutrinos},
{\sl Phys. Rev.} {\bf D 70}, 075004, (2004).
\bibitem{huong} D. T. Huong, M. C. Rodriguez and H. N. Long, 
{\it Scalar sector of supersymmetric 
$SU(3_{C}\otimes U(1)_{N}$ model with 
right-handed neutrinos},
arXiv:hep-ph/0508045.

\bibitem{oravinez} J. C. Montero, V. Pleitez and O. Ravinez,
{\it Soft superweak CP violation in a $331$ model}, 
{\sl Phys. Rev.}{\bf D 60}, 076003, (1999).
\bibitem{Montero:2005yb}J. C. Montero, C. C. Nishi, V. Pleitez, O. Ravinez and M. C. Rodriguez,
{\it Soft CP violation in $K$ meson systems},
{\sl Phys. Rev.}{\bf D73}, 016003, (2006).

\bibitem{Aad:2015zhl}G. Aad {\it et al.} [ATLAS and CMS Collaborations], {\it Combined Measurement of the Higgs Boson Mass in $pp$ Collisions at $\sqrt{s}=7$ and 8 TeV with the ATLAS and CMS Experiments},
{\sl Phys. Rev. Lett.}{\bf 114}, 191803, (2015), 
[arXiv:1503.07589 [hep-ex]].


\bibitem{tonasse}M. D. Tonasse, {\it The scalar sector of $3-3-1$ 
models}, {\sl Phys. Lett.}{\bf B381}, 191, (1996).

\bibitem{Diaz:2003dk} R. A. Diaz, R. Martinez and F. Ochoa,
{\it The Scalar sector of the 
$SU(3)_{L}\otimes U(1)_{X}$ model},
{\sl Phys. Rev.}{\bf D69}, 095009, (2004),
[arXiv:hep-ph/0309280 [hep-ph]].

\bibitem{Nguyen:1998ui}
T. A. Nguyen, N. A. Ky and H. N. Long,
{\it The Higgs sector of the minimal 3 3 1 model revisited},
{\sl Int. J. Mod. Phys.}{\bf A15}, 283, (2000), [arXiv:hep-ph/9810273 [hep-ph]].

\bibitem{Ponce:2002sg}
W. A. Ponce, Y. Giraldo and L. A. Sanchez,
{\it Minimal scalar sector of 3-3-1 models without exotic electric charges},
{\sl Phys. Rev.} {\bf D67}, 075001 (2003).

\bibitem{Giraldo:2011gd}Y. Giraldo and W. A. Ponce,
{\it Scalar Potential Without Cubic Term in 3-3-1 Models Without Exotic Electric Charges},
{\sl Eur. Phys. J.}{\bf C71}, 1693 (2011), [arXiv:1107.3260 [hep-ph]].

\bibitem{CarcamoHernandez:2014mlk}
A. E. C\'arcamo Hern\'andez, R. Martinez and J. Nisperuza,
{\it $S_3$ discrete group as a source of the quark mass and mixing pattern in $331$ models},
{\sl Eur. Phys. J.}{\bf C75}, no.2, 72, 
(2015), [arXiv:1401.0937 [hep-ph]].

\bibitem{pdg} S. Eidelman {\it et al.}, Phys. Lett. B {\bf 592} (2004).

\bibitem{DeConto:2015eia}G. De Conto, A. C. B. Machado and V. Pleitez,
{\it Minimal 3-3-1 model with a spectator sextet},
{\sl Phys. Rev.} {\bf D92}, 075031, (2015)

\end{thebibliography}
\end{document}